\documentclass[nocopyright]{cimento}
\usepackage{tabularx,amsmath,braket,amssymb,physics,graphicx}

\title{Local vs global dynamics in a dissipative qubit-impurity system}

\author{G.E.~Chiatto\from{ins:x}\ETC,
G.~Chiriac\`o\from{ins:x},
G.~Falci\from{ins:x}
\atque
E.~Paladino\from{ins:x}}
\instlist{\inst{ins:x} Dipartimento di Fisica e Astronomia ``Ettore Majorana'', Universit\`{a} di Catania, Italy}

\begin{document}

\maketitle

\begin{abstract}
We analyse the dynamics of a qubit coupled to a dissipative impurity by comparing local and global derivation schemes of a Gorini–Kossakowski–Sudarshan–Lindblad (GKSL) master equation within the Born–Markov and full secular (FS) approximations. We show that the local approach correctly captures a crossover in the dynamics of the qubit coherence, while the FS approximation restricts the validity of the global approach to regimes with well-separated energy scales. Our results clarify the domains of validity of the two approaches and show that the local scheme provides a better GKSL description of the qubit dynamics in the experimentally relevant parameter regime.
\end{abstract}

\section{Introduction}
Quantum impurities are the main source of random telegraph $1/f$ noise responsible for decoherence \cite{RevModPhysPaladino,Pel,Vac,Nalbach,lofra} in solid state quantum coherent systems \cite{kjaergaard_superconducting_2020,Chi,Gia}. Within this context, the dynamics of a qubit coupled to an incoherent two-level impurity with Markovian dissipation has been extensively studied \cite{Paladino2002,Paladino2003}. A single impurity, coupled to a bosonic or fermionic environment, may induce a nontrivial dynamics of the qubit coherence, featuring monotonic decay and oscillatory regimes \cite{Paladino2005,Paladino2008}. However, the derivation of a consistent GKSL master equation \cite{Lindblad76,gorini_completely_1976} for the composite qubit--impurity system is not unique, since only part of the system is directly coupled to the environment. Indeed, local and global derivation schemes may lead to inequivalent dynamical descriptions \cite{CattaneoEtAl, DeChiara_2018} -- e.g. the local approach leads to paradoxes~\cite{levy_local_2014,purkayastha_tunable_2020} in thermal transport~\cite{pekola_colloquium_2021}. 
The present work aims to analyse the qubit--impurity system within this local--global framework, clarifying the domains of validity of each approach and identifying which scheme provides a more complete GKSL description of the qubit dynamics.

\section{Physical model and master equations}
The qubit--impurity system (see Fig. \ref{Fig.comparison} (a)) is described by the  Hamiltonian
\begin{equation}
\label{qubit-impurity hamiltonian}
    H=-\frac{\epsilon}{2}\sigma_z -\frac{\Delta}{2}\sigma_x 
    - \frac{v}{2} \sigma_z\tau_z -\frac{\epsilon_I}{2} \tau_z ,\quad
    H_B=\sum_k \omega_k a^\dagger_k a_k , \quad 
    H_{int}=\tau_+ B + \tau_- B^\dagger.
\end{equation}
where $\sigma$ ($\tau$) are the Pauli matrices of the qubit (impurity); the qubit (impurity) energy splitting is $\Omega=\sqrt{\epsilon^2 + \Delta^2}$ ($\epsilon_I$).
The two subsystems couple via a $z$--$z$ interaction with strength $v$. The impurity is coupled to a bosonic reservoir at equilibrium temperature $\beta^{-1}$ (we set $\hbar=k_B=1$) via $H_{int}$, where $B=\sum_k T_k a_k$ and $a_k$ are the bosonic annihilation operators. The environmental modes are weakly and uniformly coupled to the system, $T_k=T\ll\epsilon_I$. In this regime, the Born--Markov approximation applies and a Markovian Bloch--Redfield equation \cite{BreuerPetruccione} is derived for the four-level system in Eq. \eqref{qubit-impurity hamiltonian} \cite{RevModPhysPaladino,Paladino2003}.

In order to obtain a master equation in GKSL form, a secular approximation must be performed.
However, since only the impurity is directly coupled to the environment, while the qubit is affected indirectly through the coherent qubit--impurity interaction, the outcome of the FS approximation depends on whether a local or a global approach is adopted.
As shown in Ref. \cite{CattaneoEtAl}, these two choices may lead to different physical predictions.
In the following, we analyse both approaches, discussing their respective regimes of validity and the dynamical features they are able to capture.

\subsection{Local approach}
In the local approach, we assume that the coupling between the qubit and the impurity in Eq.~\eqref{qubit-impurity hamiltonian} is small, namely $v\ll \Omega, \epsilon_I$. In this regime we can perform the FS approximation as if the qubit and impurity were decoupled, i.e. by considering the spectrum of Eq.~\eqref{qubit-impurity hamiltonian} when $v=0$. The resulting local jump operators are
$L_\pm^{(\text{loc})} = \mathbb{I}\otimes\tau^\pm$,
respectively associated with the rates $\gamma_+$ and $\gamma_-$, where $\gamma_- = 2\pi \mathcal{N}(\epsilon_I)|T|^2$ ($\mathcal{N}$ is the density of states of the bath) \cite{Falci2004}, and it is related to $\gamma_+$ via the fluctuation--dissipation theorem $\gamma_- = e^{\beta\epsilon_I}\gamma_+$. The master equation for the qubit--impurity system then reads
\begin{equation}
    \dot\rho(t) = -i[H,\rho(t)] 
    + \gamma_- \mathcal{D}\!\left[L_-^{(\text{loc})}\right]\rho(t)
    + \gamma_+ \mathcal{D}\!\left[L_+^{(\text{loc})}\right]\rho(t).
\end{equation}

We solve the equation in the pure-dephasing limit $\Delta=0$, where the qubit populations remain constant in time \cite{Paladino2003}. Defining $\rho^I(t) = \mathrm{Tr}_Q\,\rho(t)$ and $\rho^Q(t) = \mathrm{Tr}_I\,\rho(t)$, the coherence of the qubit evolves as
$\rho^Q_{01}(t) = \Lambda^{(\text{L})}(t)\rho_{01}(0)$, with
\begin{equation}
\label{Local Lambda}
    \Lambda^{(\text{L})}(t) = e^{i\epsilon t}
    \left[
    A e^{-\frac{\gamma}{2}(1-\alpha)t}
    + (1-A)e^{-\frac{\gamma}{2}(1+\alpha)t}
    \right].
\end{equation}
where
$A = \left[(1+\alpha) + i g \delta p_0\right]/(2\alpha)$,
$\alpha = \sqrt{1+2ig \delta \bar{p} - g^2}$,
$\delta \bar{p} = (\gamma_- - \gamma_+)/\gamma$,
$\gamma=\gamma_-+\gamma_+$,
$\delta \bar{p}_0 = \rho^I_{00}(0) - \rho^I_{11}(0)$,
and $g = 2v/\gamma$. From now on we set $\delta\bar p = 0$.

The parameter $g$ tunes two different decoherence regimes for the qubit. For $g<1$, $\alpha$ is purely real and the modulus of Eq.~\eqref{Local Lambda} decays monotonically, following an exponential law. Conversely, for $g>1$, $\alpha$ becomes imaginary and the modulus of Eq.~\eqref{Local Lambda} is $|\Lambda^{(\text{L})}(t)| =\left[\cos\!\left( \frac{\gamma\delta}{2} t \right)+\frac{1}{\delta}\sin\!\left( \frac{\gamma\delta}{2} t \right)\right]e^{-\frac{\gamma t}{2}}$,
implying that the coherence does not decay monotonically in time and displays revivals. Therefore, the parameter $g$ tunes a qualitative change in the dynamics of the system.

\subsection{Global approach}
In the global approach, we perform the FS approximation on the total spectrum of Eq.~\eqref{qubit-impurity hamiltonian}, which in our case can be considered fully non--degenerate when $|\Omega_0-\Omega_1|>\gamma$, $\Omega_\tau = \sqrt{(\epsilon+v(1-2\tau))^2+\Delta^2}$ and $\tau=0,1$ denotes the excitation state of the impurity. When this condition is satisfied, the FS approximation is valid and a GKSL master equation can be derived. Conversely, relaxing this condition does not necessarily lead to a GKSL equation, for which physical properties such as complete positivity must be explicitly verified \cite{Paladino2003}.
 
The decay jump operators now are
$L^{(\text{g})}_{1,2} = c\,\ketbra{\pm_0,0}{\pm_1,1}$,
both associated with the decay rate $\gamma_-$, where $\{\ket{\pm_\tau,\tau}\}$ are the eigenstates of $H$ and
$c=\cos[(\theta_0-\theta_1)/2]$, $\theta_\tau = \tan^{-1}[\Delta/(\epsilon+v(1-2\tau))]$. The corresponding absorption jump operators are $(L^{(\text{g})}_{1,2})^\dagger$ with rate $\gamma_+$.
The resulting GKSL master equation for the qubit--impurity system is

\begin{equation}
    \dot\rho(t) = -i[H,\rho(t)] +\left(\gamma_- \mathcal{D}\left[ L_1^{(\text{g})} \right] 
      + \gamma_- \mathcal{D}\left[ L_2^{(\text{g})} \right] +\gamma_+ \mathcal{D}\left[ L_1^{(\text{g})\dagger} \right] 
      + \gamma_+ \mathcal{D}\left[ L_2^{(\text{g})\dagger} \right]\right) \rho(t).
\end{equation}

Analogously to the local approach, we analyse the dynamics in the limit $\Delta=0$ where $|\Omega_0-\Omega_1|>\gamma$ is equivalent to $2v>\gamma$, namely $g>1$. In this case, the qubit coherence evolves as
$\rho^Q_{01}(t) = \Lambda^{(\text{G})}(t)\rho^Q_{01}(0)$, with $\Lambda^{(\text{G})}(t) = e^{i\epsilon t}\left[e^{ivt}e^{-\gamma_+ t}+ e^{-ivt}e^{-\gamma_- t}\right]$. In the infinite-temperature limit, one finds
$|\Lambda^{(\text{G})}(t)| = 2e^{-\gamma t/2}|\cos vt|$,
showing that the coherence exhibits revivals with pseudo-period $T=\pi/v$. In conclusion, the global approach is structurally incapable of providing a complete description of the crossover around $g=1$ within the FS approximation.

\begin{figure}
  \centering
  \includegraphics[width=\textwidth]{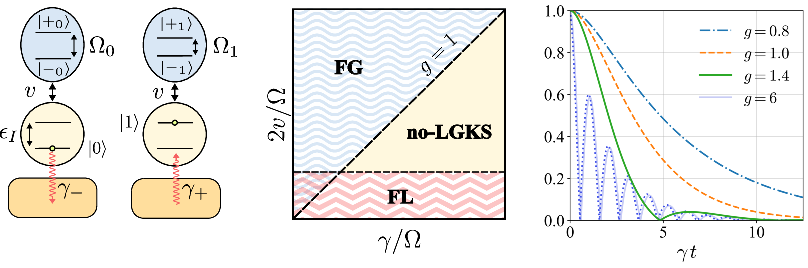}
  \put(-285,130){(a)}
  \put(-157,130){(b)}
  \put(-75,130){$|\Lambda(t)|$}
  \put(-15,130){(c)}
  \put(-242,35){$\times$}
  \caption{(a) Sketch of the qubit (blue) coupled to an incoherent impurity (yellow), in turn coupled to a bath (orange). The impurity undergoes emission (absorption) processes with rate $\gamma_-$ ($\gamma_+$) and the coupling $v$ induces a conditional shift of the qubit energy splitting.
(b) Diagram in the $(v,\gamma)$ plane showing the regions where a GKSL master equation can be derived within the FS approximation. The blue/red regions correspond to the global (FG)/ local (FL) approach; in the yellow region no GKSL description is assured. In the region marked by ``$\times$'' both approaches are valid.
(c) Modulus of the qubit coherence in the FL approach, showing the crossover from monotonic decay ($g<1$) to dynamics with revivals ($g>1$). The dotted curve shows the FG solution for $g=6$, which converges to the FL solution. Parameters: $\epsilon=20$, $\Delta=0$, $\epsilon_I=20$, $\delta\bar p=\delta p_0=0$.
}
  \label{Fig.comparison}
\end{figure}
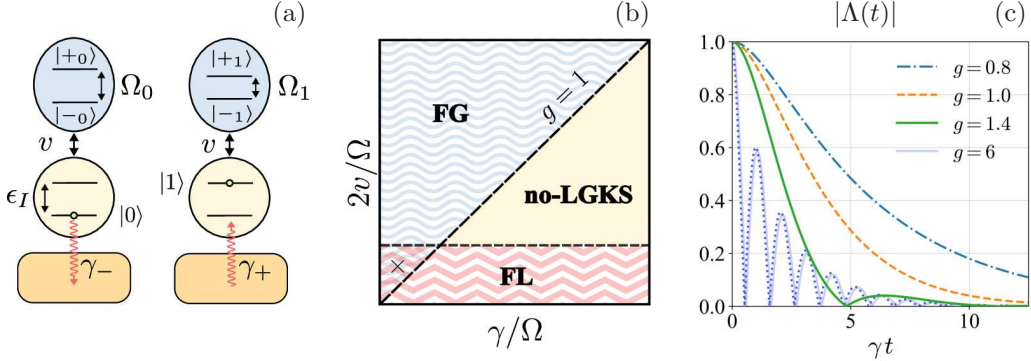

\subsection*{Comparison between the two approaches} 

From the results obtained, we deduce that the local approach is able to reproduce all the results observed in other previous studies \cite{Paladino2002,Paladino2003}, as opposed to the FS global approach. Indeed, the validity of the two methods crucially depends on the interaction regime of the system. The diagram shown in Fig. \ref{Fig.comparison} (b) summarises the domains of validity of each approach in the parameter space $(v,\gamma)$. In particular, the local approach relies on the perturbative condition $v\ll\Omega,\epsilon_I$ and allows for a GKSL description independently of the ratio $2v/\gamma$. Conversely, the global approach requires a non-degenerate energy spectrum, expressed by the condition $|\Omega_0-\Omega_1|>\gamma$. Outside these regions, the FS approximation cannot be reliably applied. 

The different validity domains influence the dynamical information accessible via the two methods. The local approach captures a crossover at $g=1$ between monotonic decay of the qubit coherence and a revival regime displaying damped oscillations. In contrast, the global approach only describes oscillatory dynamics, see indigo line in Fig.~\ref{Fig.comparison}(c), and coincides with the oscillatory behaviour of the local solution for $g>1$ (blue dotted line). The monotonic decay regime is therefore structurally inaccessible within the global approach. Importantly, this is not related to the strength of the qubit--impurity coupling itself, but to the rigid application of the FS approximation on the eigenstates of the total system Hamiltonian. 
The validity of the global approach is extended to the non-secular approach of Ref.~\cite{Paladino2003}, where the qubit decay is neglected, yielding an analytical quantum map which can be extended to pulsed dynamics~\cite{Falci2004} analogous to the 
partial secular approximation~\cite{CattaneoEtAl,vaaranta_numerical_2026}. The global non-secular approach  generally leads to non-GKSL dynamics, which lies outside the scope of the present work.

\acknowledgments
G. Chiatto and G. Chiriacò are supported by ICSC, project E63C22001000006. E.P. and G. F. acknowledge the PNRR MUR Project PE0000023-NQSTI and the UniCT-Piaceri 2024-26 QTCM project. GF acknowledges support from PRIN 2022 ``SuperNISQ".

\bibliographystyle{varenna}
\bibliography{bibliography}

\end{document}